\documentclass[11pt,a4paper,fleqn]{article}

\usepackage[tbtags]{amsmath}
\usepackage{amssymb}
\usepackage{mathrsfs}

\newcommand{\ud}{\mathrm{d}}

\newcommand{\DB}{\mathrm{DB}}
\newcommand{\SB}{\mathrm{SB}}

\title{\textbf{Symmetries of Snyder--de Sitter space and relativistic particle dynamics}}

\author{{\normalsize \textbf{Rabin Banerjee$^{1}$, Kuldeep Kumar$^{2}$ and Dibakar Roychowdhury$^{1}$}}\\[0.5ex]
{\small $^1$\textit{S.N. Bose National Centre for Basic Sciences,}}\\
{\small \textit{JD Block, Sector III, Salt Lake, Kolkata 700098, India}}\\[0.2ex]
{\small $^2$\textit{Department of Physics, Panjab University, Chandigarh 160014, India}}\\[0.2ex]
{\small \texttt{rabin@bose.res.in}, \texttt{kuldeepk@pu.ac.in}, \texttt{dibakar@bose.res.in}}
}

\date{}

\begin{document}

\maketitle

\begin{abstract}
We study the deformed conformal-Poincar\'e symmetries consistent with the Snyder--de Sitter space. A relativistic particle model invariant under these deformed symmetries is given. This model is used to provide a gauge independent derivation of the Snyder--de Sitter algebra. Our results are valid in the leading order in the parameters appearing in the model.
\end{abstract}



\section{\label{sec:intro}Introduction}

The Snyder model \cite{Snyder:1946qz}, introduced long time ago in 1947, showed the possibility of defining a noncommutative model of spacetime without breaking the Lorentz invariance. The noncommutativity emerges from the Poisson brackets of the position coordinates that involve a constant $\kappa$ which can be identified with the Planck energy. In the limit $\kappa \to \infty$ one recovers the standard phase space of special relativity.

More recently, the extension of the Snyder model to a de Sitter background has evoked much interest \cite{KowalskiGlikman:2004kp, Mignemi:2008fj, Mignemi:2009zz, Carrisi:2010jv, Guo:2004p, Guo:2007sf}. This model now involves, besides the speed of light, two other observer-independent constants, the Planck energy and the de Sitter radius which is related to the cosmological constant. Consequently, it is also known as doubly special relativity in de Sitter space or triply special relativity \cite{KowalskiGlikman:2004kp, Mignemi:2008fj}. Several interesting features of this model have been discussed in the papers \cite{KowalskiGlikman:2004kp, Mignemi:2008fj, Mignemi:2009zz, Carrisi:2010jv}.

The Snyder--de Sitter algebra has been obtained by an appropriate gauge fixing of a dynamical model defined in six dimensions \cite{Mignemi:2009zz}. Also, it has been derived using methods of two-time physics \cite{Carrisi:2010jv}.

One of the motivations of the present paper is to provide a gauge independent derivation of the Snyder--de Sitter algebra from a dynamical model in usual one-time physics. Apart from the fact that such a derivation is lacking in the literature, it provides a deeper insight into the model. Also, since such a model for the Snyder case was found by one of us, in a collaborative work \cite{Banerjee:2006wf}, the treatment of the Snyder--de Sitter example ought to be a logical extension of the methods developed in \cite{Banerjee:2006wf}. We show that the dynamical model highlights the role of the conformal-Poincar\'e symmetries. We work out the complete set of deformed conformal-Poincar\'e generators and demonstrate their closure. The modified transformation laws are also shown to be compatible with the Snyder--de Sitter algebra. The connection of the results in the Snyder--de Sitter space with the standard space is illuminated by an explicit map relating our dynamical model with the usual relativistic free-particle model. Apart from Sec.~\ref{sec:sds} all our results are valid in the leading order in the parameters---Planck energy $\kappa$, which is related to the noncommutativity (Snyder) parameter, and de Sitter radius $\alpha$, which is related to the cosmological constant---appearing in the Snyder--de Sitter space.

This paper is organised as follows. After a brief introduction to the Snyder--de Sitter algebra in Sec.~\ref{sec:sds}, we construct a dynamical model which leads to the Snyder--de Sitter algebra in Sec.~\ref{sec:model}. In Sec.~\ref{sec:mapping} we provide maps relating the variables in the usual (commutative) space and those in the Snyder--de Sitter space. These maps, which preserve the fundamental Poisson brackets of the usual space, are used to obtain the generators and the transformations compatible with the Snyder--de Sitter algebra. In Sec.~\ref{sec:noether} we carry out a Noether analysis of our model and precisely reproduce those conformal-Poincar\'e generators found earlier in Sec.~\ref{sec:mapping} using certain maps. Concluding remarks are left for Sec.~\ref{sec:conlu}.


\section{\label{sec:sds}Snyder--de Sitter space}

Although the generalisation of the Snyder model to a de Sitter background is not unique \cite{Mignemi:2008fj}, an elegant extension has been proposed in \cite{KowalskiGlikman:2004kp}. This classical Snyder--de Sitter algebra is given by
\begin{equation}\label{sds-alg}
\begin{aligned}
&\!\{ X_\mu, X_\nu \} = - \frac{1}{\kappa^2} ( X_\mu P_\nu - X_\nu P_\mu ), \\
&\!\{ P_\mu, P_\nu \} = - \frac{1}{\alpha^2} ( X_\mu P_\nu - X_\nu P_\mu ), \\
&\!\{ X_\mu, P_\nu \} = \eta_{\mu \nu} - \frac{1}{\alpha^2} X_\mu X_\nu - \frac{1}{\kappa^2} P_\mu P_\nu - \frac{2}{\alpha \kappa} P_\mu X_\nu,
\end{aligned}
\end{equation}
where $\kappa$ is the Planck energy and $\alpha$ is the de Sitter radius. In the limit $\kappa \to \infty$ one recovers the algebra of a free particle in de Sitter space, in the limit $\alpha \to \infty$ one recovers the Snyder algebra and when $\alpha$ and $\kappa$ both tend to infinity one recovers the usual algebra. The Snyder--de Sitter algebra given in \eqref{sds-alg} is a classical algebra and satisfies various Jacobi identities.\footnote{This algebra cannot be elevated straightforwardly to the quantum level since Jacobi identities do not hold in that case, implying the need of appropriate ordering. It may be compared with the Snyder algebra where operator ordering ambiguities do not exist \cite{Banerjee:2006wf}.}

Just like the Snyder algebra, the algebra \eqref{sds-alg} is compatible with the standard Lorentz symmetry. If we consider the infinitesimal Lorentz transformation,
\begin{gather}
\label{deltaX} \delta X_\mu = \omega_{\mu \rho} X^\rho,\\
\label{deltaP} \delta P_\mu = \omega_{\mu \rho} P^\rho,
\end{gather}
with $\omega_{\mu \rho}$ antisymmetric, then the variation on the left-hand side of the $X$--$X$ bracket yields
\begin{equation}
\begin{split}
\delta \{ X_\mu, X_\nu \} &= \{ \delta X_\mu, X_\nu \} + \{ X_\mu, \delta X_\nu \} \\
&= - \frac{1}{\kappa^2} {\omega_\mu}^\rho ( X_\rho P_\nu - X_\nu P_\rho ) + \frac{1}{\kappa^2} {\omega_\nu}^\rho ( X_\rho P_\mu - X_\mu P_\rho ),
\end{split}
\end{equation}
which is also obtained if we take the variation on the right-hand side of the $X$--$X$ bracket and then use \eqref{deltaX} and \eqref{deltaP}. An identical treatment follows for the $P$--$P$ and $X$--$P$ brackets given in \eqref{sds-alg}. Expectedly, the primitive form of the Lorentz generator,
\begin{equation}
J_{\mu \nu} = X_\mu P_\nu - X_\nu P_\mu,
\end{equation}
yields the correct transformation:
\begin{gather}
\delta X_\mu = - \tfrac{1}{2} \omega^{\rho \sigma} \{ X_\mu, J_{\rho \sigma} \} = \omega_{\mu \rho} X^\rho,\\
\delta P_\mu = - \tfrac{1}{2} \omega^{\rho \sigma} \{ P_\mu, J_{\rho \sigma} \} = \omega_{\mu \rho} P^\rho.
\end{gather}

The explicit presence of phase-space variables in the algebra, however, hints at a possible deformation of the translation symmetry. One can easily check that the standard transformation law for translation, $\delta X_\mu = a_\mu$, $\delta P_\mu = 0$, is not compatible with \eqref{sds-alg}. So the transformation rule must be modified to achieve consistency.

We start with the general expressions for $\delta X_\mu$ and $\delta P_\mu$ which are dimensionally consistent and demand the compatibility with \eqref{sds-alg}. This finally yields the deformed translation as\footnote{Notation: $A \cdot B = A_\mu B^\mu$, $A^2 = A \cdot A = A_\mu A^\mu$.}
\begin{gather}
\label{deltaX-trans2} \delta X_\mu = a_\mu - \frac{1}{\alpha^2} (a \cdot X) X_\mu - \frac{1}{\kappa^2} (a \cdot P) P_\mu - \frac{2}{\alpha \kappa} (a \cdot X) P_\mu,\\
\label{deltaP-trans2} \delta P_\mu = \frac{1}{\alpha^2} \left[ (a \cdot X) P_\mu - (a \cdot P) X_\mu \right].
\end{gather}
Although we have a deformed transformation rule for translations, the generator retains its primitive form
\begin{equation}
T_\mu = P_\mu
\end{equation}
so that
\begin{gather}
\delta X_\mu = a^\rho \{ X_\mu, T_\rho \},\\
\delta P_\mu = a^\rho \{ P_\mu, T_\rho \}
\end{gather}
reproduce transformations \eqref{deltaX-trans2} and \eqref{deltaP-trans2}, respectively. However, because of the nonvanishing $P$--$P$ bracket, the generators satisfy a modified Poincar\'e algebra:
\begin{equation}\label{poincare-d}
\begin{aligned}
&\!\{ T_\mu, T_\nu \} = - \frac{1}{\alpha^2} J_{\mu \nu}, \\
&\!\{ J_{\mu \nu}, T_\rho \} = \eta_{\mu \rho} T_\nu - \eta_{\nu \rho} T_\mu, \\
&\!\{ J_{\mu \nu}, J_{\rho \sigma} \} = \eta_{\mu \rho} J_{\nu \sigma} - \eta_{\nu \rho} J_{\mu \sigma} - \eta_{\mu \sigma} J_{\nu \rho} + \eta_{\nu \sigma} J_{\mu \rho}.
\end{aligned}
\end{equation}
This algebra also follows from the following considerations. First note that $J_{\mu \nu}$ are the generators of Lorentz algebra $\mathrm{SO}(1,3)$. Now for a de Sitter space, which can be realised as a hyperboloid embedded in 5-dimensional flat space, the full symmetry will be $\mathrm{SO}(1,4)$. This includes the old operators $J_{\mu \nu}$ and $T_0$ ($T_i$), which effect the velocity (space) transformations of the old coordinates to the new ones. Expectedly, for $\alpha \to \infty$, corresponding results \cite{Banerjee:2006wf} for the Snyder algebra are reproduced.

Now we look for a possible extension to the conformal sector. Taking the primitive form of the dilatation generator,
\begin{equation}
D = X \cdot P,
\end{equation}
we find that the $D$--$P$ bracket does not close:
\begin{equation}
\{ D, T_\mu \} = T_\mu - \frac{1}{\kappa^2} P^2 P_\mu - \frac{1}{\alpha^2} X^2 P_\mu - \frac{2}{\alpha \kappa} P^2 X_\mu.
\end{equation}
Therefore, we will have to look for some appropriate modification of the generator and the transformation so as to achieve consistency with \eqref{sds-alg} for the full conformal group. This will, in general, lead to deformed generators even in the Poincar\'e sector.


\section{\label{sec:model}Dynamical model leading to Snyder--de Sitter algebra}

We want to construct a model that leads to the Snyder--de Sitter algebra. For that we take input from the corresponding dynamical model for the Snyder space discussed in \cite{Banerjee:2006wf}. But there is an important point which needs to be elaborated here. Since the $P$--$P$ bracket does not change in the Snyder case, the dispersion relation also does not change from that in the usual space. In the de Sitter space, however, the $P$--$P$ bracket changes and so does the dispersion relation. Contrary to the Minkowski space, de Sitter space does not admit a natural choice of coordinates and therefore different quantities have simpler expressions in different coordinate systems, such as natural coordinates, conformal coordinates and Beltrami coordinates \cite{Mignemi:2008fj}. For natural coordinates, the metric induced on the de Sitter hyperboloid by the 5-dimensional flat metric reads
\begin{equation}\label{metric-dS}
g_{\mu \nu} = \eta_{\mu \nu} - \frac{2}{\alpha^2}\, \frac{X_\mu X_\nu}{1 + 2 X^2/\alpha^2}, \quad g^{\mu \nu} = \eta^{\mu \nu} + \frac{2}{\alpha^2} X^\mu X^\nu.
\end{equation}
Now the Lagrangian for a massive relativistic particle in arbitrary space with metric $g_{\mu \nu}$ is
\begin{equation}
L = \frac{1}{2} \left( \frac{1}{\eta} g_{\mu \nu} \dot{X}^\mu \dot{X}^\nu + \eta m^2 \right),
\end{equation}
and the canonical Hamiltonian is
\begin{equation}
H_\mathrm{C} = \frac{\eta}{2} \left( g^{\mu \nu} P_\mu P_\nu - m^2 \right),
\end{equation}
where $P_\mu$ is the conjugate momentum. If we take the $g^{\mu \nu}$ given in \eqref{metric-dS} then the canonical Hamiltonian becomes
\begin{equation}\label{input1}
H_\mathrm{C} = \frac{\eta}{2} \left[ P^2 - m^2 + \frac{2}{\alpha^2} (X^\mu P_\mu)^2 \right],
\end{equation}
clearly revealing the modification in the usual dispersion relation.

Now we recall the action for the dynamical model discussed in \cite{Banerjee:2006wf} which yields the Snyder algebra:
\begin{equation}\label{input2}
S = \int \ud \tau \left[ - X \cdot \dot{P} - \frac{1}{\kappa^2} (X \cdot P) (P \cdot \dot{P}) - e ( P^2 - m^2 ) \right],
\end{equation}
where $e$ is a Lagrange multiplier enforcing the Einstein condition $P^2 - m^2 = 0$ and we have kept the terms up to the leading order only. With the two inputs, \eqref{input1} and \eqref{input2}, we write down the action for our dynamical model as
\begin{equation}\label{action}
\begin{split}
S = \int \ud \tau \Big[ 
 &- X \cdot \dot{P} - \frac{1}{\kappa^2} (X \cdot P) (P \cdot \dot{P}) - \frac{1}{\alpha^2} \Big\{ X^2 (\dot{X} \cdot P + X \cdot \dot{P}) + (X \cdot P) (X \cdot \dot{X}) \Big\} \\
 &- \frac{1}{\alpha \kappa} X^2 (P \cdot \dot{P}) - e \Big\{ P^2 - m^2 + \frac{2}{\alpha^2} (X \cdot P)^2 \Big\}
\Big],
\end{split}
\end{equation}
which contains additional dimensionally consistent terms and  goes over to the action \eqref{input2} in the limit $\alpha \to \infty$. Note that the dispersion relation in the Snyder--de Sitter space is the same as that of the de Sitter space and our model corresponds to the natural parametrisation of the de Sitter coordinates.

\subsection{Dirac's constraint analysis}

We shall now perform a constraint analysis of the dynamical model computing the Dirac brackets which will yield the Snyder--de Sitter algebra. We interpret $X$ and $P$ of the action \eqref{action} as the configuration-space variables in an extended space. The canonical momenta conjugate to $X$, $P$ and $e$ are
\begin{gather}
\Pi^X_\mu = \frac{\partial L}{\partial \dot{X}^\mu} = -\frac{1}{\alpha^2} \left[ X^2 P_\mu + (X \cdot P) X_\mu \right], \\
\Pi^P_\mu = \frac{\partial L}{\partial \dot{P}^\mu} = - X_\mu - \frac{1}{\kappa^2} (X \cdot P) P_\mu - \frac{1}{\alpha^2} X^2 X_\mu - \frac{1}{\alpha\kappa} X^2 P_\mu, \\
\Pi^e = \frac{\partial L}{\partial \dot{e}} = 0. 
\end{gather}
Since none of these momenta involve velocities, these are, following Dirac \cite{Dirac:1964}, primary constraints of the theory:
\begin{gather}
\label{phi}
\phi = \Pi^e \approx 0, \\
\label{phi1}
\phi_{1,\mu} = \Pi^X_\mu + \frac{1}{\alpha^2} \left[ X^2 P_\mu + (X \cdot P) X_\mu \right] \approx 0, \\
\label{phi2}
\phi_{2,\mu} = \Pi^P_\mu + X_\mu + \frac{1}{\kappa^2} (X \cdot P) P_\mu + \frac{1}{\alpha^2} X^2 X_\mu + \frac{1}{\alpha\kappa} X^2 P_\mu \approx 0.
\end{gather}
The Poisson algebra of the constraints is given by\footnote{The only nonvanishing brackets here are $\{X_\mu, \Pi^X_\nu\} = \{P_\mu, \Pi^P_\nu\} = \eta_{\mu\nu}$ and $\{e, \Pi^e\} = 1$. Also, we restrict ourselves to the leading order in the parameters, retaining only terms up to $1/\alpha^2$, $1/\kappa^2$ and $1/\alpha \kappa$ and ignoring higher-order terms. This approximation is implicit in the following results unless stated otherwise.}
\begin{equation}
\label{consalg}
\begin{aligned}
\!&\{ \phi, \phi \} = \{ \phi, \phi_{1,\mu} \} = \{ \phi, \phi_{2,\mu} \} =0, \\
\!&\{ \phi_{1,\mu}, \phi_{1,\nu} \} = - \frac{1}{\alpha^2} ( X_\mu P_\nu - X_\nu P_\mu ), \\
\!&\{ \phi_{2,\mu}, \phi_{2,\nu} \} = - \frac{1}{\kappa^2} ( X_\mu P_\nu - X_\nu P_\mu ), \\
\!&\{ \phi_{1,\mu}, \phi_{2,\nu} \} = - \left( \eta_{\mu \nu} + \frac{1}{\kappa^2} P_\mu P_\nu + \frac{1}{\alpha^2} X_\mu X_\nu + \frac{2}{\alpha \kappa} X_\mu P_\nu \right).
\end{aligned}
\end{equation}
The constraint $\phi$ is first-class as it has vanishing brackets with all the constraints. The constraints $\phi_{1,\mu}$ and $\phi_{2,\mu}$ are second-class and will be eliminated by the use of Dirac brackets. The canonical Hamiltonian is\footnote{Although our analysis is restricted to the leading order in the parameters, this expression for the canonical Hamiltonian is an exact result.}
\begin{equation}
H_\mathrm{C} = e \left[ P^2 - m^2 + \frac{2}{\alpha^2} (X \cdot P)^2 \right]
\end{equation}
and therefore the total Hamiltonian is written as
\begin{equation}
H_\mathrm{T} = e \left[ P^2 - m^2 + \frac{2}{\alpha^2} (X \cdot P)^2 \right] + \lambda \phi + \lambda_1^\mu \phi_{1,\mu} + \lambda_2^\mu \phi_{2,\mu}.
\end{equation}
Time consistency of constraint $\phi$ leads to the secondary constraint
\begin{equation}
\label{psi}
\psi = \{\phi, H_\mathrm{T}\} = P^2 - m^2 + \frac{2}{\alpha^2} (X \cdot P)^2 \approx 0.
\end{equation}

The second-class constraints $\phi_{1,\mu}$ and $\phi_{2,\mu}$ will now be eliminated using Dirac brackets. For that we compute the constraint matrix:
\begin{equation}\label{cmatix}
\begin{split}
\Lambda_{\mu\nu}
&=
\begin{pmatrix}
\{ \phi_{1,\mu}, \phi_{1,\nu} \} & \{ \phi_{1,\mu}, \phi_{2,\nu} \} \\[1.0ex]
\{ \phi_{2,\mu}, \phi_{1,\nu} \} & \{ \phi_{2,\mu}, \phi_{2,\nu} \}
\end{pmatrix}
\\
&=
\begin{pmatrix}
- \frac{1}{\alpha^2} (X_\mu P_\nu - X_\nu P_\mu) & - \eta_{\mu\nu} - \big( \frac{P_\mu P_\nu}{\kappa^2} + \frac{X_\mu X_\nu}{\alpha^2} + \frac{2 X_\mu P_\nu}{\alpha \kappa} \big) \\[1.0ex]
\eta_{\mu\nu} + \big( \frac{P_\mu P_\nu}{\kappa^2} + \frac{X_\mu X_\nu}{\alpha^2} + \frac{2 P_\mu X_\nu}{\alpha \kappa} \big) & - \frac{1}{\kappa^2} (X_\mu P_\nu - X_\nu P_\mu)
\end{pmatrix},
\end{split}
\end{equation}
which has the inverse
\begin{equation}\label{cmatix-inv}
(\Lambda^{-1})^{\mu\nu} =
\begin{pmatrix}
- \frac{1}{\kappa^2} (X^\mu P^\nu - X^\nu P^\mu) & \eta^{\mu\nu} - \big( \frac{P^\mu P^\nu}{\kappa^2} + \frac{X^\mu X^\nu}{\alpha^2} + \frac{2 P^\mu X^\nu}{\alpha \kappa} \big) \\[1.0ex]
- \eta^{\mu\nu} + \big( \frac{P^\mu P^\nu}{\kappa^2} + \frac{X^\mu X^\nu}{\alpha^2} + \frac{2 X^\mu P^\nu}{\alpha \kappa} \big) & - \frac{1}{\alpha^2} (X^\mu P^\nu - X^\nu P^\mu)
\end{pmatrix},
\end{equation}
such that $(\Lambda^{-1})_{ij}^{\mu \nu} \Lambda_{jk,\nu \rho} = \delta_{ik} \delta^\mu_\rho$ ($i,j = 1,2$). Now using the definition,
\begin{equation}
\{f, g\}_{\DB} = \{f, g\} - \{f, \phi_{i,\mu}\} (\Lambda^{-1})_{ij}^{\mu \nu} \{\phi_{j,\nu}, g\},
\end{equation}
the relevant Dirac brackets among the configuration-space variables for our model are
\begin{equation}
\label{DB}
\begin{aligned}
&\!\{ X_\mu, X_\nu \}_\DB = - \frac{1}{\kappa^2} ( X_\mu P_\nu - X_\nu P_\mu ), \\
&\!\{ P_\mu, P_\nu \}_\DB = - \frac{1}{\alpha^2} ( X_\mu P_\nu - X_\nu P_\mu ), \\
&\!\{ X_\mu, P_\nu \}_\DB = \eta_{\mu \nu} - \frac{1}{\alpha^2} X_\mu X_\nu - \frac{1}{\kappa^2} P_\mu P_\nu - \frac{2}{\alpha \kappa} P_\mu X_\nu.
\end{aligned}
\end{equation}
This algebra is basically the Snyder--de Sitter algebra \eqref{sds-alg}.

The secondary constraint $\psi$ now has vanishing Dirac brackets with all other constraints and is therefore first-class. Thus our model has two first-class constraints $\phi$ and $\psi$ and two second-class constraints $\phi_{1,\mu}$ and $\phi_{2,\mu}$, which have been eliminated using the Dirac brackets. The first-class constraint $\phi$ is not a meaningful constraint as it is just the momentum corresponding to the Lagrange multiplier $e$ appearing in \eqref{action}. The other first-class constraint $\psi$ is a generator of gauge transformation. It should have vanishing or weakly vanishing Dirac brackets with the conformal-Poincare generators since these are physical variables and hence gauge invariant. We will demonstrate this fact later in Sec.~\ref{sec:mapping} where the explicit form of these generators has been derived. 

\subsection{Symplectic analysis}

There is an alternative method of getting the basic brackets that does not require explicit classification of the constraints. This is the symplectic approach \cite{sm1974, Faddeev:1988qp} and is geared for first-order systems. Since \eqref{action} is such a system let us apply this approach here.

The Euler--Lagrange equations of motion for $X$ and $P$ following from the action \eqref{action} are
\begin{gather}
\label{EL1}
\begin{split}
\dot{P}_\mu &+ \frac{1}{\kappa^2} (P \cdot \dot{P}) P_\mu \\ {} &+ \frac{1}{\alpha^2} \left[ (\dot{X} \cdot P) X_\mu + (X \cdot \dot{P}) X_\mu - (X \cdot \dot{X}) P_\mu + 4 e (X \cdot P) P_\mu \right] + \frac{2}{\alpha \kappa} (P \cdot \dot{P}) X_\mu = 0,
\end{split} \\
\label{EL2}
\begin{split}
\dot{X}_\mu &- 2 e P_\mu + \frac{1}{\kappa^2} \left[ (\dot{X} \cdot P) P_\mu + (X \cdot \dot{P}) P_\mu - (P \cdot \dot{P}) X_\mu \right] \\ {} &+ \frac{1}{\alpha^2} \left[ (X \cdot \dot{X}) X_\mu - 4 e (X \cdot P) X_\mu \right] + \frac{2}{\alpha \kappa} (X \cdot \dot{X}) P_\mu = 0
\end{split}.
\end{gather}
Equations \eqref{EL1} and \eqref{EL2} can be written as
\begin{equation}
\Lambda_{ij,\mu\nu} \dot{\xi}_j^\nu = \partial_{i,\mu} H_\mathrm{C}
\end{equation}
in the notation
\begin{equation}
\xi_1^\mu = X^\mu, \quad \xi_2^\mu = P^\mu, \quad \partial_{i,\mu} = \frac{\partial}{\partial \xi_i^\mu},
\end{equation}
while $\Lambda$ is given in \eqref{cmatix}.

Now we compute the symplectic brackets, which are defined as
\begin{equation}
\{f, g\}_{\SB} = (\Lambda^{-1})_{ij}^{\mu \nu} \partial_{i,\mu}f \, \partial_{j,\nu}g.
\end{equation}
The relevant brackets for our model turn out to be
\begin{equation}
\label{SB}
\begin{aligned}
&\!\{ X_\mu, X_\nu \}_\SB = - \frac{1}{\kappa^2} ( X_\mu P_\nu - X_\nu P_\mu ), \\
&\!\{ P_\mu, P_\nu \}_\SB = - \frac{1}{\alpha^2} ( X_\mu P_\nu - X_\nu P_\mu ), \\
&\!\{ X_\mu, P_\nu \}_\SB = \eta_{\mu \nu} - \frac{1}{\alpha^2} X_\mu X_\nu - \frac{1}{\kappa^2} P_\mu P_\nu - \frac{2}{\alpha \kappa} P_\mu X_\nu,
\end{aligned}
\end{equation}
which are identical with the Dirac brackets \eqref{DB} and the Snyder--de Sitter algebra \eqref{sds-alg}.


\section{\label{sec:mapping}Deformed symmetries in Snyder--de Sitter space}

Now we discuss algebraic transformations mapping the usual canonical $(x, p)$ to the $(X, P)$ of Snyder--de Sitter space. This mapping is obtained by comparing the action \eqref{action} to that for a free relativistic particle in the usual space,
\begin{equation}
S = \int \ud \tau \left[ - x \cdot \dot{p} - e ( p^2 - m^2 ) \right],
\end{equation}
where $\{ x_\mu, x_\nu \} = \{ p_\mu, p_\nu \}$ = 0, $\{ x_\mu, p_\nu \} = \eta_{\mu \nu}$. These two actions are mapped by the transformations
\begin{gather}
\label{mapx}
x_\mu = X_\mu + \frac{1}{\kappa^2} (X \cdot P) P_\mu + \frac{1}{\alpha \kappa} X^2 P_\mu, \\
\label{mapp}
p_\mu = P_\mu + \frac{1}{\alpha^2} (X \cdot P) X_\mu.
\end{gather}
These maps have the inverse
\begin{gather}
\label{mapx-inv}
X_\mu = x_\mu - \frac{1}{\kappa^2} (x \cdot p) p_\mu - \frac{1}{\alpha \kappa} x^2 p_\mu, \\
\label{mapp-inv}
P_\mu = p_\mu - \frac{1}{\alpha^2} (x \cdot p) x_\mu.
\end{gather}
Now we can obtain the generators and the transformations for the full conformal-Poincar\'e sector. Taking the generators in the usual space and applying the maps \eqref{mapx} and \eqref{mapp} yields the generators in the Snyder--de Sitter space:
\begin{equation}
G = G(x(X,P),p(X,P)).
\end{equation}
The final form of the generators $T_\mu$ (translation), $J_{\mu \nu}$ (Lorentz transformation), $D$ (dilatation) and $K_\mu$ (special conformal transformation) thus obtained is
\begin{gather}
\label{gen-T}
T_\mu = P_\mu + \frac{1}{\alpha^2} (X \cdot P) X_\mu, \\
\label{gen-J}
J_{\mu \nu} = X_\mu P_\nu - X_\nu P_\mu, \\
\label{gen-D}
D = X \cdot P + \frac{1}{\alpha^2} (X \cdot P) X^2 + \frac{1}{\kappa^2} (X \cdot P) P^2 + \frac{1}{\alpha \kappa} X^2 P^2, \\
\label{gen-K}
K_\mu = 2 (X \cdot P) X_\mu - X^2 P_\mu + \frac{2}{\kappa^2} (X \cdot P) P^2 X_\mu + \frac{1}{\alpha^2} (X \cdot P) X^2 X_\mu + \frac{2}{\alpha \kappa} X^2 P^2 X_\mu.
\end{gather}
The Lorentz generator retains its primitive form while other generators are deformed.

Various transformations are now obtained using these generators. For translation we get
\begin{gather}
\label{deltaX-trans}
\delta X_\mu = a^\rho \{ X_\mu, T_\rho \} = a_\mu - \frac{1}{\kappa^2} (a \cdot P) P_\mu - \frac{2}{\alpha \kappa} (a \cdot X) P_\mu, \\
\label{deltaP-trans}
\delta P_\mu = a^\rho \{ P_\mu, T_\rho \} = - \frac{1}{\alpha^2} \left[ (a \cdot P) X_\mu + (X \cdot P) a_\mu \right].
\end{gather}
Therefore the transformation rule for the translation is deformed. Lorentz transformation, like the generator $J_{\mu \nu}$, is however not deformed:
\begin{gather}
\label{deltaX-Lorentz}
\delta X_\mu = -\tfrac{1}{2} \omega^{\rho \sigma} \{ X_\mu, J_{\rho \sigma} \} = \omega_{\mu \rho} X^\rho, \\
\label{deltaP-Lorentz}
\delta P_\mu = -\tfrac{1}{2} \omega^{\rho \sigma} \{ P_\mu, J_{\rho \sigma} \} = \omega_{\mu \rho} P^\rho.
\end{gather}
Transformation rule for the dilatation,
\begin{gather}
\label{deltaX-dia}
\delta X_\mu = \epsilon \{ X_\mu, D \} = \epsilon \left[ X_\mu + \frac{2}{\kappa^2} (X \cdot P) P_\mu \right], \\
\label{deltaP-dia}
\delta P_\mu = \epsilon \{ P_\mu, D \} = -\epsilon \left[ P_\mu + \frac{2}{\alpha^2} (X \cdot P) X_\mu \right],
\end{gather}
is deformed and same is the case with the special conformal transformation:
\begin{gather}
\label{deltaX-sc}
\begin{split}
\delta X_\mu
 &= \omega^\rho \{ X_\mu, K_\rho \} \\
 &= 2 (\omega \cdot X) X_\mu - X^2 \omega_\mu \\
 &{} \quad\, + \frac{1}{\kappa^2} \left[ 6(\omega \cdot X) (X \cdot P) P_\mu - (\omega \cdot P) X^2 P_\mu \right] + \frac{2}{\alpha \kappa} (\omega \cdot X) X^2 P_\mu,
\end{split}\\
\label{deltaP-sc}
\begin{split}
\delta P_\mu
 &= \omega^\rho \{ P_\mu, K_\rho \} \\
 &= -2 (X \cdot P) \omega_\mu + 2 (\omega \cdot P) X_\mu - 2 (\omega \cdot X) P_\mu \\
 &{} \quad\, - \frac{2}{\kappa^2} (X \cdot P) P^2 \omega_\mu - \frac{1}{\alpha^2} \left[ (X \cdot P) X^2 \omega_\mu + (\omega \cdot P) X^2 X_\mu \right] - \frac{2}{\alpha \kappa} X^2 P^2 \omega_\mu.
\end{split}
\end{gather}
All these transformations correctly reproduce the usual ones in the limit $\alpha \to \infty$, $\kappa \to \infty$.

We now show that these transformations are consistent with the Snyder--de Sitter algebra \eqref{sds-alg}. Let us consider the $X$--$X$ bracket of \eqref{sds-alg} and the special conformal transformation, \eqref{deltaX-sc} and \eqref{deltaP-sc}. We compute the variation on the left-hand side of the $X$--$X$ bracket:
\begin{equation}
\begin{split}
\delta \{ X_\mu, X_\nu \} &= \{ \delta X_\mu, X_\nu \} + \{ X_\mu, \delta X_\nu \} \\
 &= \frac{1}{\kappa^2} \left[ \omega_\mu \left\{ X^2 P_\nu -2 (X \cdot P) X_\nu \right\} - \omega_\nu \left\{ X^2 P_\mu -2 (X \cdot P) X_\mu \right\} \right],
\end{split}
\end{equation}
where we have used \eqref{deltaX-sc} and the algebra \eqref{sds-alg} to write down the second step. The same expression is obtained by taking the variation on the right-hand side of the $X$--$X$ bracket,
\begin{equation}
\delta \left[- \frac{1}{\kappa^2} ( X_\mu P_\nu - X_\nu P_\mu ) \right] = - \frac{1}{\kappa^2} ( \delta X_\mu P_\nu + X_\mu \delta P_\nu - \delta X_\nu P_\mu - X_\nu \delta P_\mu ),
\end{equation}
and using \eqref{deltaX-sc} and \eqref{deltaP-sc}. Similarly, we find that for the $P$--$P$ and $X$--$P$ brackets in \eqref{sds-alg} the variation on the left-hand side yields the same expression as the variation on the right-hand side under the transformations \eqref{deltaX-sc} and \eqref{deltaP-sc}. This shows the consistency of the special conformal transformation with the algebra \eqref{sds-alg}. A similar exercise can be done for other transformations.

Expectedly, the generators, which do not retain their primitive form except $J_{\mu\nu}$, satisfy the usual conformal-Poincar\'e algebra:
\begin{equation}\label{algebra}
\begin{aligned}
&\!\{ T_{\mu}, T_{\nu} \} = 0, \quad
 \{ T_{\mu}, J_{\rho \sigma} \} = - \eta_{\mu \rho} T_{\sigma} + \eta_{\mu \sigma} T_{\rho}, \\
&\!\{ J_{\mu \nu}, J_{\rho \sigma} \} = \eta_{\mu \rho} J_{\nu \sigma} - \eta_{\nu \rho} J_{\mu \sigma} - \eta_{\mu \sigma} J_{\nu \rho} + \eta_{\nu \sigma} J_{\mu \rho}, \\
&\!\{ D, T_{\mu} \} = T_{\mu}, \quad
 \{ D, J_{\mu \nu} \}  = 0, \quad
 \{ D, D \}  = 0, \quad
 \{ K_{\rho}, T_{\mu} \} = 2 \left( \eta_{\rho \mu} D + J_{\rho \mu} \right), \\
&\!\{ K_{\rho}, J_{\mu \nu} \} = - \eta_{\rho \mu} K_{\nu} + \eta_{\rho \nu} K_{\mu}, \quad
 \{ K_{\rho}, D \} = K_{\rho}, \quad
 \{ K_{\rho}, K_{\mu} \} = 0.
\end{aligned}
\end{equation}
Note that the Poincar\'e sector of this algebra, which is the usual Poincar\'e algebra, is different from the modified Poincar\'e algebra \eqref{poincare-d}, which has a nonvanishing $T$--$T$ bracket.

Finally, we show the gauge invariance of the conformal-Poincar\'e generators by computing their algebra with the first-class constraint $\psi$ \eqref{psi}. The brackets of $\psi$ with $T_\mu$ and $J_{\mu \nu}$ vanish strongly,
\begin{equation}
\{ T_\mu , \psi \}_\DB = \{ J_{\mu \nu} , \psi \}_\DB = 0.
\end{equation}
For the dilatation generator we find
\begin{equation}
\{ D, \psi \}_\DB = 2 \left[ P^2 + \frac{2}{\alpha^2} (X \cdot P)^2 \right],
\end{equation}
where the right-hand side involves the constraint $\psi$ \eqref{psi} for the massless version and hence is weakly zero. For the special conformal generator we have
\begin{equation}\label{splcnf}
\{ K_\mu, \psi \}_\DB = 4 \left[ P^2 + \frac{2}{\alpha^2} (X \cdot P)^2 \right] X_\mu + 4 P^2 \left[ \frac{X \cdot P}{\kappa^2} + \frac{X^2}{\alpha \kappa} \right] P_\mu.
\end{equation}
Here again the first term on the right-hand side involves the constraint $\psi$ for the massless case and if we use this constraint \eqref{psi} to replace $P^2$ appearing in the second term by $-2 (X \cdot P)^2/\alpha^2$ then this term also drops out, since we are restricting to the leading order in the parameters. The brackets of $\psi$ with $D$ and $K_\mu$ therefore vanish weakly. 


\section{\label{sec:noether}Noether's theorem and generators}

Now that we have the various transformation laws we will carry out a Noether analysis of \eqref{action} to study the invariance of the action and to reproduce the generators. This will also serve as a consistency check for our results. Invariance of an action under an infinitesimal symmetry transformation,
\begin{equation}
\delta q^\mu = \{ q^\mu, G \},
\end{equation}
is given by
\begin{equation}
\label{deltaS-Noeth}
\delta S = \delta \int \ud \tau L(q, \dot{q}) = \int \ud \tau \frac{\ud B }{\ud \tau},
\end{equation}
where $B = p_\mu \delta q^\mu - G$, $p_\mu$ being the canonical momentum conjugate to $q^\mu$. For our model \eqref{action} both $X$ and $P$ are interpreted as configuration-space variables and so the generator is
\begin{equation}
G = \Pi^X_\mu \delta X^\mu + \Pi^P_\mu \delta P^\mu - B,
\end{equation}
which, after implementing the constraints \eqref{phi1} and \eqref{phi2}, reads
\begin{equation}\label{gen-Noeth}
\begin{split}
G = &- \frac{1}{\alpha^2} \left[ X^2 P_\mu + (X \cdot P) X_\mu \right] \delta X^\mu \\
    &- \left[ X_\mu + \frac{1}{\kappa^2} (X \cdot P) P_\mu + \frac{1}{\alpha^2} X^2 X_\mu + \frac{1}{\alpha \kappa} X^2 P_\mu \right] \delta P^\mu - B.
\end{split}
\end{equation}

Expectedly, under Lorentz transformation, the action \eqref{action} is invariant ($\delta S = 0$), therefore \eqref{deltaS-Noeth} implies $B = 0$. Using \eqref{deltaX-Lorentz} and \eqref{deltaP-Lorentz}, \eqref{gen-Noeth} yields
\begin{equation}
G = - \tfrac{1}{2} \omega^{\mu \nu} \left( X_\mu P_\nu - X_\nu P_\mu \right)  = - \tfrac{1}{2} \omega^{\mu \nu} J_{\mu \nu}
\end{equation}
reproducing $J_{\mu \nu}$ \eqref{gen-J}. Under (deformed) translation, \eqref{deltaX-trans} and \eqref{deltaP-trans}, it is also invariant:
\begin{equation}
\label{deltaS-trans}
\delta S = \int \ud \tau \frac{\ud}{\ud \tau} \left[ -\left( a \cdot P + \frac{1}{\alpha^2} (a \cdot X) (X \cdot P) \right) \right],
\end{equation}
as the integrand is a total derivative. Now comparison of \eqref{deltaS-trans} with \eqref{deltaS-Noeth} reveals that
\begin{equation}
B = - \left[ a \cdot P + \frac{1}{\alpha^2} (a \cdot X) (X \cdot P) \right].
\end{equation}
Using \eqref{deltaX-trans} and \eqref{deltaP-trans} in \eqref{gen-Noeth} we get
\begin{equation}
G = - B = a \cdot P + \frac{1}{\alpha^2} (a \cdot X) (X \cdot P) = a^\mu T_\mu,
\end{equation}
reproducing the translation generator $T_\mu$ \eqref{gen-T}.

Variation of the action \eqref{action} under dilatation, \eqref{deltaX-dia} and \eqref{deltaP-dia}, is given by
\begin{equation}
\delta S = \int \ud \tau \left[ 2 \epsilon e \left( P^2 + \frac{2}{\alpha^2} (X \cdot P)^2 \right) \right].
\end{equation}
Here the integrand cannot be expressed as a total derivative. However, if we pass to the constraint shell \eqref{psi} then invariance is achieved for the massless version. Now $B = 0$ and using \eqref{deltaX-dia} and \eqref{deltaP-dia} in \eqref{gen-Noeth} we get $G = \epsilon D$, where $D$ is given by \eqref{gen-D}. Finally, for the special conformal transformation, \eqref{deltaX-sc} and \eqref{deltaP-sc}, variation of the action gives
\begin{equation}\label{deltaS-K}
\begin{split}
\delta S = \int \ud \tau \Big[& \frac{\ud}{\ud \tau}\Big\{ 2 (\omega \cdot X) (X \cdot P) - (\omega \cdot P) X^2 + \frac{2}{\kappa^2} (\omega \cdot X) (X \cdot P) P^2 \\ &\qquad {} + \frac{1}{\alpha^2} (\omega \cdot X) (X \cdot P) X^2 + \frac{2}{\alpha \kappa} (\omega \cdot X) X^2 P^2 \Big\} \\
& {} + 4e (\omega \cdot X) \Big( P^2 + \frac{2}{\alpha^2} (X \cdot P)^2 \Big) + 4e (\omega \cdot P) P^2 \Big( \frac{X \cdot P}{\kappa^2} + \frac{X^2}{\alpha \kappa} \Big) \Big].
\end{split}
\end{equation}
Again, for the massless version if we pass to the constraint shell \eqref{psi} then the last two terms in the integrand drop out by precisely following the logic discussed below \eqref{splcnf}, and \eqref{deltaS-K} simplifies to
\begin{equation}
\delta S = \int \ud \tau \Big[ \frac{\ud}{\ud \tau} ( \omega \cdot K ) \Big],
\end{equation}
where $K$ is given by \eqref{gen-K}. Hence $B = \omega \cdot K$ and using \eqref{deltaX-sc} and \eqref{deltaP-sc}, \eqref{gen-Noeth} yields
\begin{equation}
G = 2 ( \omega \cdot K ) - B = \omega \cdot K = \omega^\mu K_\mu,
\end{equation}
reproducing the special conformal generator. Conformal symmetries are broken for massive theories, as happens in the usual space.

Thus it is possible to reproduce the generators from a Noether analysis. This also provides a link with the earlier way of obtaining the generators using suitable maps connecting conventional phase-space algebra and the Snyder--de Sitter algebra.


\section{\label{sec:conlu}Conclusions}

We have made an exhaustive study of the deformed conformal-Poincar\'e symmetries consistent with the Snyder--de Sitter space. This is the first such study in this model. Contrary to the analysis done for canonical (constant) noncommutativity \cite{Banerjee:2006db}, here the results are valid in the leading order in the parameters $\kappa$ (related to the noncommutativity (Snyder) parameter) and $\alpha$ (related to the cosmological constant).  A relativistic particle model invariant under these deformed symmetries was given. This model was then used to provide a gauge independent derivation of the Snyder--de Sitter algebra. Also, our model was directly defined in four dimensions. In this context we note that dynamical models which reproduce the Snyder--de Sitter type algebra \cite{Mignemi:2009zz, Carrisi:2010jv, Romero:2004er} are generally constructed in higher dimensions and a gauge fixing is required to reduce to four dimensions. Both these issues were bypassed here. The difficulty in constructing a dynamical model directly in four dimensions is essentially connected to the appearance of two scales ($\alpha$ and $\kappa$) rather than only one (either $\alpha$ or $\kappa$). Indeed an earlier failed attempt in this direction \cite{Jaroszkiewicz:1995} highlights this issue.

We gave a mapping between the usual phase space and the Snyder--de Sitter space. Exploiting this map, the full set of conformal-Poincar\'e generators of our dynamical model was constructed. These deformed generators were used to explicitly derive the deformed transformations. Although the generators were deformed, they satisfied the standard conformal-Poincar\'e algebra.

The Noether analysis of the model was performed. The deformed spacetime transformations were now used to compute the relevant generators. It was reassuring to note that the results agreed with those obtained algebraically by enforcing the various maps.


\section*{Acknowledgements}

KK would like to thank S.N. Bose National Centre for Basic Sciences for hospitality where this work was carried out. DR thanks Council of Scientific and Industrial Research (CSIR), Government of India, for financial support.




\begin{thebibliography}{99}\raggedright \small \setlength{\itemsep}{0.0cm}

\bibitem{Snyder:1946qz} H.S. Snyder, ``Quantized space-time,'' {\em Phys.~Rev.} 71 (1947) 38.

\bibitem{KowalskiGlikman:2004kp} J. Kowalski-Glikman and L. Smolin, ``Triply special relativity,'' {\em Phys.\ Rev.} D 70 (2004) 065020.

\bibitem{Mignemi:2008fj} S. Mignemi, ``Doubly special relativity in de Sitter spacetime,'' {\em Annalen Phys.} 522 (2010) 924.

\bibitem{Mignemi:2009zz} S. Mignemi, ``The Snyder--de Sitter model from six dimensions,'' {\em Class.\ Quant.\ Grav.} 26 (2009) 245020.

\bibitem{Carrisi:2010jv} M.C. Carrisi and S. Mignemi, ``Snyder--de Sitter model from two-time physics,'' arXiv:1010.6258 [gr-qc].

\bibitem{Guo:2004p} H.-Y. Guo, C.-G. Huang, Z. Xu and B. Zhou, ``On special relativity with cosmological constant,'' {\em Phys.\ Lett.} A 331 (2004) 1.

\bibitem{Guo:2007sf} H.-Y. Guo, C.-G. Huang, Y. Tian, H.-T. Wu and B. Zhou, ``Snyder's model---de Sitter special relativity duality and de Sitter gravity,'' {\em Class.\ Quant.\ Grav.} 24 (2007) 4009.

\bibitem{Banerjee:2006wf} R. Banerjee, S. Kulkarni and S. Samanta, ``Deformed symmetry in Snyder space and relativistic particle dynamics,'' {\em JHEP} 0605 (2006) 077.

\bibitem{Dirac:1964} P.A.M. Dirac, {\em Lectures on Quantum Mechanics}, Belfer Graduate School of Science, Yeshiva University, New York, 1964.

\bibitem{sm1974} E.C.G. Sudarshan and N. Mukunda, {\em Classical Dynamics: A Modern Perspective}, John Wiley \& Sons Inc., 1974.

\bibitem{Faddeev:1988qp} L.D. Faddeev and R. Jackiw, ``Hamiltonian reduction of unconstrained and constrained systems,'' {\em Phys.~Rev.~Lett.} 60 (1988) 1692.

\bibitem{Banerjee:2006db} R. Banerjee and K. Kumar, ``Deformed relativistic and nonrelativistic symmetries on canonical noncommutative spaces,'' {\em Phys.~Rev.} D 75 (2007) 045008.

\bibitem{Romero:2004er} J.M. Romero and A. Zamora, ``Snyder noncommutative space-time from two-time physics,'' {\em Phys.\ Rev.} D 70 (2004) 105006.

\bibitem{Jaroszkiewicz:1995} G. Jaroszkiewicz, ``A dynamical model for the origin of Snyder's quantized spacetime algebra,'' {\em J.~Phys.~A: Math.\ Gen.} 28 (1995) L343.

\end{thebibliography}
\end{document}